\documentclass[aps,prl,showpacs,twocolumn,superscriptaddress]{revtex4}
\usepackage{amsmath, amssymb}
\usepackage{braket}
\usepackage{ulem}
\usepackage{soul}
\usepackage{dcolumn}
\usepackage{bm}
\usepackage{graphicx}
\usepackage{wasysym}
\usepackage{setspace}
\usepackage{mathrsfs}
\usepackage{color}
\usepackage{hyperref}
\usepackage{CJKutf8}
\hypersetup{colorlinks=true,linkcolor=blue,filecolor=blue,citecolor = blue,urlcolor=blue,}

\usepackage{tikz}
\usepackage{lipsum}
\begin{document}
\title{Skyrmion Hall effect and shape deformation of current-driven bilayer skyrmions in synthetic antiferromagnets}
\author{Mu-Kun Lee}
\affiliation
{Department of Applied Physics, Waseda University, Okubo, Shinjuku-ku, Tokyo 169-8555, Japan}
\author{Javier A. V\'{e}lez}
\affiliation{Donostia International Physics Center, 20018 San Sebasti\'{a}n, Spain}
\affiliation{Polymers and Advanced Materials Department: Physics, Chemistry, and Technology, University of the Basque Country, UPV/EHU, 20018 San Sebasti\'{a}n, Spain}
\author{Rub\'{e}n M. Otxoa}
\affiliation{Hitachi Cambridge Laboratory, J. J. Thomson Avenue, Cambridge CB3 0HE, United Kingdom}
\author{Masahito Mochizuki}
\affiliation
{Department of Applied Physics, Waseda University, Okubo, Shinjuku-ku, Tokyo 169-8555, Japan}
\begin{abstract}
The commonly believed absence of skyrmion Hall effect for topologically trivial magnetic skyrmions is reconsidered for bilayer skyrmions in synthetic antiferromagnets driven by spin-transfer and spin-orbit torques. Using a general Lagrangian formalism, we show that Bloch-type bilayer skyrmions acquire a finite Hall angle when driven by spin-orbit torque, while N\'{e}el-type skyrmions do not, in agreement with micromagnetic simulations. Both types of skyrmions exhibit current-induced elliptical deformation with minor and major axes aligned longitudinally and transversely to their velocity, respectively. A linear relation between velocity and longitudinal radius is derived with a coefficient proportional to the strength of spin-orbit torque. These effects are critical for antiferromagnetic skyrmion-based applications such as skyrmion racetrack memory. The Lagrange equations also reproduce the linear Hall-angle--helicity relation reported by~\href{https://journals.aps.org/prapplied/abstract/10.1103/PhysRevApplied.17.064015}{Msiska \textit{et al.}, Phys. Rev. Appl. \textbf{17}, 064015 (2022)}. An intuitive explanation of the skyrmion Hall effect for arbitrary helicity based on the antiferromagnetic exchange torque is also provided.
\end{abstract}
\maketitle	
\section{I. Introduction}
Magnetic skyrmions~\cite{Nagaosa,Seki_Mochizuki,Everschor_rev} are whirl-like textures of local magnetic moments characterized by a nontrivial topological number representing how many times the local moments wrap a sphere~\cite{Braun}. Skyrmions in ferromagnets have been theoretically predicted~\cite{Bogdanov01,Bogdanov02,Rossler} and later experimentally observed in a lattice form in chiral magnets~\cite{Muhlbauer2009,YuXZ2010}. Owing to the topological protection, skyrmions are robust against thermal excitations and can be stabilized at room temperatures~\cite{YuXZ2011}. They can even be driven into motion by a current density five or six orders of magnitude smaller than that required to move the domain walls~\cite{SkHE00,Iwasaki01,Iwasaki02}. Therefore, they are promising candidates for magnetic devices like the racetrack memory~\cite{Fert}. 

However, when being driven by the current-induced spin-transfer torque (STT)~\cite{Slonczewski,Zhang_Li}, the skyrmions suffer from the skyrmion Hall effect (SkHE) that induces a transverse velocity component relative to the direction of applied current, as a result of the Magnus force due to the finite topological number~\cite{SkHE00,SkHE01,SkHE02}. The SkHE sets an obstacle for applications of, e.g., skyrmion racetrack memory~\cite{SkHE01} that requires skyrmions to move along the current direction. Several methods to reduce SkHE have been proposed~\cite{SkHE_reduced01,SkHE_reduced02,SkHE_reduced03,SkHE_reduced04,SkHE_reduced05}. One common belief is that the topologically trivial magnetic textures such as antiferromagnetic (AF) skyrmions~\cite{AF_sky01,AF_sky02,AF_sky03,AF_sky04} can be free from SkHE due to the cancellation of opposite Magnus forces acting on the individual skyrmions in the two magnetic sublattices. 
Surprisingly, Msiska \textit{el al.}~\cite{Msiska} recently adopted the Thiele equation~\cite{Thiele} to calculate the motion of topologically trivial textures including a pair of bilayer skyrmions in synthetic antiferromagnets (AFMs)~\cite{Yang_Parkin,Yang_Parkin_2017,AF_sky04} and a skyrmionium~\cite{skyrmionium} driven by current-induced spin-orbit torque (SOT)~\cite{SOT00,SOT01}. An unexpected finite Hall angle is found for the Bloch skyrmion, while it is vanishing for the N\'{e}el skyrmion. The Hall angle follows a linear dependence on the helicity of skyrmion, in the case with only SOT but without STT.

Since a skyrmion is authentically a two-dimensional texture in magnetic thin films, the internal structural degrees of freedom may correlate with its center-of-mass motion. However, despite few works~\cite{Chen,Liu} have been devoted to the generalizations to deformed skyrmions without structural circular symmetry, most theoretical works using Thiele equation presume a rigidly rotational-symmetric texture during motion which is indistinctive to its static profile, and this approach is hard to incorporate the effect on skyrmion velocity by their shape deformation, or vice versa. 
In the case of current-driven motion of AF domain walls (DWs), for example, it is crucial to consider the Lorentz contraction of DW width which results in a nonlinear current dependence of the DW velocity~\cite{Shiino,LeeDW}, which is drastically different from the linear behavior in ferromagnets. The Lorentz invariance is ubiquitous in the dynamics of staggered magnetization in AFMs when the current-induced torques and damping terms which break the Lorentz symmetry are compensated or small. In a result, for AF skyrmions the contracted (elongated) skyrmion radius in the direction parallel (perpendicular) to the current direction has also been found by solving the equations of motion within the Lagrangian framework by Salimath \textit{et al.}~\cite{Salimath}. In their work, however, only N\'{e}el skyrmions have been considered and no skyrmion Hall effect has been found. Therefore, a comprehensive theory based on Lagrangian formalism to consider both STT and SOT for topologically trivial deformed skyrmions with various helicities is desirable and important for the potential applications of skyrmions in AF spintronics.

In this paper, we first derive the Landau-Lifshitz-Gilbert-Slonczewski (LLGS) equation for the staggered magnetization in a synthetic AFM under STT and SOT, and find the corresponding Lagrangian and Rayleigh dissipation, which are in general applicable to any magnetic textures in the same type of system and under the same torques. We then solve the Lagrange equation for a pair of antiferromagnetically coupled skyrmions (the so-called bilayer skyrmions~\cite{AF_sky01}) to obtain their velocity and elliptical shape deformation as functions of the skyrmion helicity and applied current. A finite (zero) skyrmion Hall angle is obtained for the Bloch (N\'{e}el) bilayer skyrmions in the presence of damping-like SOT.
Our theory reduces exactly to the linear Hall-angle--helicity relation in~\cite{Msiska} when the nonadiabatic STT is ignored. However, as a generalization of the Thiele approach by Lagrangian formalism, we find a robust linear dependence between the skyrmion velocity and longitudinal radius along its moving direction with the factor being proportional to the strength of SOT, thus our result may provide a way for experiments to measure the SOT magnitude in synthetic AFMs. Micromagnetic simulations are in good agreement with our Lagrangian theory of skyrmion velocity and the ratio between the lengths of major and minor axes in an elliptical ansatz of the skyrmion profile. We also provide an intuitive view by analyzing the AF exchange torque~\cite{Yang_Parkin,Yang_Parkin_2017} driven by SOT to understand the occurrence of finite (zero) SkHE for Bloch (N\'{e}el) bilayer skyrmions, and the Hall angle for general cases with any helicity when both bulk and interfacial Dzyaloshinskii-Moriya interactions (DMIs) exist.
\begin{figure*}
	\centering
	\includegraphics[scale=0.4]{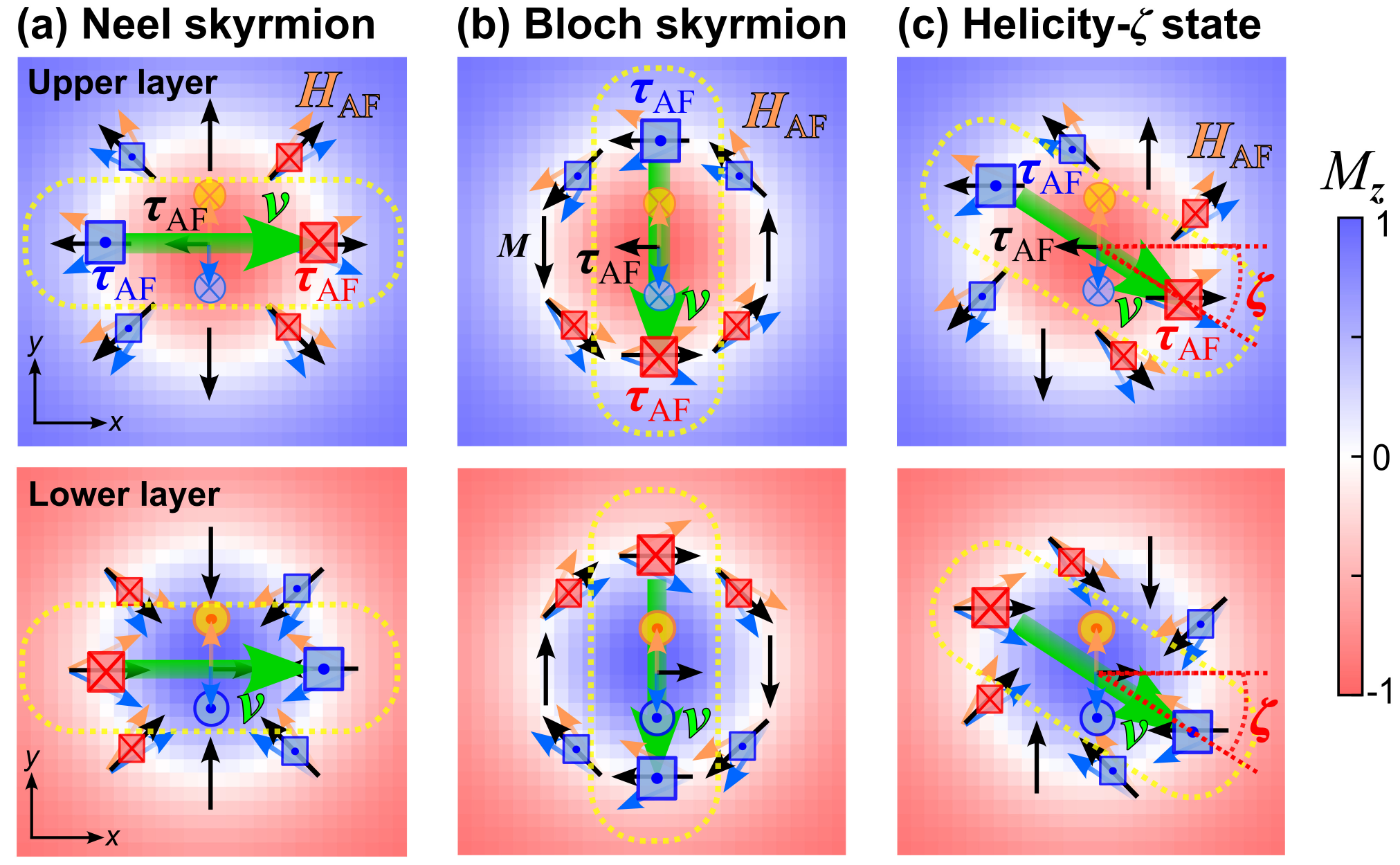}
	\caption{Schematics of AF exchange torque induced by SOT for (a) N\'{e}el,  (b) Bloch, and (c) helicity-$\zeta$ bilayer skyrmions. The upper/lower panel indicates the upper/lower magnetic layer in the synthetic AFM. Black arrows at the skyrmion peripheries denote the in-plane magnetizations in the static state, which will be schematically tilted into the blue arrows when driven by damping-like SOT. These tilted moments then induce the AF exchange field $\mathbf{H}_{\rm AF}$ in the opposite layer with directions shown by orange arrows, resulting in the exchange torque $\bm{\tau}_{\rm AF}$ with directions denoted by squared dots and crosses (with the size of squares indicating the magnitude). At skyrmion centers, blue (orange) arrows combined with blue (orange) circles indicate the tilted moments (exchange fields), and $\bm{\tau}_{\rm AF}$ is shown by black arrows. Green arrows point in the expected velocity ($\bm{v}$) directions of the skyrmions if we focus on the quasi-one-dimensional DW-like profile inside the yellow dotted boxes.}
	\label{Fig_intuitive}
\end{figure*}
\section{II. Method and result}
\subsection{II.1 Lagrangian formalism}
We first give a perspective on the form of the LLGS equation that we use to derive the Lagrangian formalism in this work. The original LLGS equation~\cite{SkHE00} is
\begin{eqnarray}
&&\dot{\mathbf{M}}_j=\gamma \mathbf{H}^{\rm eff}_j\times\mathbf{M}_j+\mathbf{T}_j+\alpha\mathbf{M}_j\times \dot{\mathbf{M}}_j\label{LLGS00},
\end{eqnarray}
where $\mathbf{M}_j$ is the unit vector pointing in the direction of local magnetization density (with saturated magnitude $M_{\rm s}$) for the two sublattices $j=$ A or B, $\gamma$ is the gyromagnetic ratio, $\mathbf{H}^{\rm eff}_j=\frac{-1}{M_{\rm s}a^3_0}\frac{\delta W}{\delta\mathbf{M}_j}$ is the effective field derived from the system energy $W$ with $a_0$ being the lattice constant, and $\mathbf{T}_j$ includes all the torques induced by the electric current. The last term is the Gilbert damping with a dimensionless constant $\alpha$. To solve the LLGS equation by numerical integration methods such as Runge-Kutta algorithms, in most cases we need to put all time derivatives of $\mathbf{M}_j$ on one side of the equation. This can be done by taking a curl of $\mathbf{M}_j$ on both sides of Eq.~(\ref{LLGS00}) to get
\begin{eqnarray}
\dot{\mathbf{M}}_j&=&\frac{1}{1+\alpha^2}\Big[\Big(\gamma\mathbf{H}^{\rm eff}_j\times\mathbf{M}_j+\mathbf{T}_j\Big)\nonumber\\
&&+\alpha\mathbf{M}_j\times\Big(\gamma\mathbf{H}^{\rm eff}_j\times\mathbf{M}_j+\mathbf{T}_j\Big)\Big].\label{LLGS01}
\end{eqnarray}
For specific forms of the torques in $\mathbf{T}_j$, the Gilbert damping essentially \textit{renormalizes} their magnitudes. In particular, we consider a synthetic AFM composed of two ferromagnetic layers lying on the $xy$ plane, which are stacked along the $z$ direction and are coupled by AF exchange interaction. Applying a current along $x$ direction in both layers, we consider the torque
\begin{eqnarray}
{\mathbf{T}}_j&=&-u \mathbf{M}'_j+\beta u\mathbf{M}_j\times\mathbf{M}'_j\nonumber\\
&+&uc_{\rm H}\mathbf{M}_j\times \mathbf{M}_j\times \hat{\mathbf{y}}+uc_{\rm H}f_{\rm so}\mathbf{M}_j\times\hat{\mathbf{y}}\label{T},
\end{eqnarray}
which includes sequentially the adiabatic and nonadiabatic STTs, and dampling-like and field-like SOTs. Here $\mathbf{M}'_j\equiv\partial_x\mathbf{M}_j$, $u=\frac{p a^3_0}{2e}J$ is the spin drift velocity, $J$ is applied current density, $p$ is electron spin polarization, and $\beta$ is dimensionless nonadiabatic STT parameter.  The strength of SOT is defined as $c_{\rm H}=\frac{\gamma \hbar \theta_{\rm sh}}{M_{\rm s}t_{\rm L} p a^3_0}$ with $\theta_{\rm sh}$ and $t_{\rm L}$ being respectively the spin Hall angle and layer thickness parameter, $\hbar$ is the reduced Planck constant, and $f_{\rm so}$ is the ratio of field-like to damping-like SOTs. 
A typical synthetic AFM~\cite{Yang_Parkin,Yang_Parkin_2017,AF_sky04} considered in this work consists of antiferromagnetically coupled ferromagnetic layers on top of a heavy-metal layer to inject the vertical spin current which exerts SOT on these two ferromagnetic layers. In general, due to the finite spin coherence length of the spin current, the strength of SOT might differ between the two magnetic layers. In this work, however, we assume equivalent magnitudes of SOT in both magnetic layers for simplicity. This assumption is justified when the magnetic layers are thinner than the spin coherent length. Then, by plugging the formula of the torque into Eq.(\ref{LLGS01}), we obtain,
\begin{eqnarray}
&&\dot{\mathbf{M}}_j=\gamma\tilde{\mathbf{H}}^{\rm eff}_j\times\mathbf{M}_j-\alpha\gamma\mathbf{M}_j\times\mathbf{M}_j\times\tilde{\mathbf{H}}^{\rm eff}_j-\tilde{u}\mathbf{M}'_j\label{LLGS1}\\
&&+\tilde{\beta}\tilde{u}\mathbf{M}_j\times\mathbf{M}'_j+\tilde{u}\tilde{c}_{\rm H}\mathbf{M}_j\times\mathbf{M}_j\times\hat{\mathbf{y}}+\tilde{u}\tilde{c}_{\rm H}\tilde{f}_{\rm so}\mathbf{M}_j\times{\hat{\mathbf{y}}}.\nonumber
\end{eqnarray}
Here the tilded parameters are defined as
\begin{eqnarray}
&&\tilde{u}=\frac{u(1+\alpha\beta)}{1+\alpha^2},\ \tilde{\beta}=\frac{\beta-\alpha}{1+\alpha\beta},\ \tilde{\mathbf{H}}^{\rm eff}_j=\frac{\mathbf{H}^{\rm eff}_j}{1+\alpha^2},\nonumber\\ 
&&\tilde{c}_{\rm H}=\frac{c_{\rm H}(1+\alpha f_{\rm so})}{1+\alpha\beta},\ \tilde{f}_{\rm so}=\frac{f_{\rm so}-\alpha}{1+\alpha f_{\rm so}}.
\end{eqnarray}
We have written the torques in the same forms as in Eq.~(\ref{T}), such that it is clear the Gilbert damping renormalizes them to give the effective tilded parameters, which are the actual strengths of torques acting on magnetizations. 
For instance, the effective dampling-like SOT as the fifth term on the right-hand side of Eq.~(\ref{LLGS1}) comes not only from the original third term in Eq.~(\ref{T}) but also from the field-like SOT, which induces another contribution proportional to $\alpha f_{\rm so}$ in $\tilde{c}_{\rm H}$. The reversed statement is also true as can be seen by the definition of $\tilde{f}_{\rm so}$. Therefore, a field-like SOT induces a dampling-like SOT, and vice versa, via the Gilbert damping. In this sense there is no pure damping-like nor pure field-like SOTs, unless the Gilbert damping is absent.
In literature, although micromagnetic simulations are mostly done by solving the renormalized LLGS equation [Eq.~(\ref{LLGS1})], analytically the Thiele equation~\cite{Msiska,Chen,Liu} or Lagrangian formalism~\cite{Salimath,Gomonay2010} are often derived from the original LLGS equation [Eq.~(\ref{LLGS00})]. However, in the later approach we find it is somehow elusive to get these correct effective parameters, unless the Gilbert damping term in Eq.~(\ref{LLGS00}) is treated properly. To make the formalism more transparent and precise, we adopt an alternative way to derive the Lagrangian formalism starting from Eq.~(\ref{LLGS1}). In this procedure, the effective tilded parameters will show up naturally.

We define the averaged and staggered magnetization vectors as $\mathbf{m}=(\mathbf{M}_{\rm A}+\mathbf{M}_{\rm B})/2$ and $\mathbf{l}=(\mathbf{M}_{\rm A}-\mathbf{M}_{\rm B})/2$, respectively, with constraints $\mathbf{m}^2+\mathbf{l}^2=1$ and $\mathbf{m}\cdot \mathbf{l}=0$ valid when the temperature is well below the Curie temperature. In the exchange limit when the interlayer AF exchange coupling strength $J_{\rm AF}(>0)$ is the largest energy scale, we have $|\mathbf{m}|\ll |\mathbf{l}|$ and $\mathbf{m}$ can be approximated by a slave variable dependent solely on the dynamics of $\mathbf{l}$, as shown in Appendix A. We can then derive an equation of motion for $\mathbf{l}$ as
\begin{eqnarray}
&&\mathcal{D}^2\mathbf{l}\approx 2(\tilde{\mathbf{H}}_f\times\mathcal{D}\mathbf{l})+(\mathcal{D}\tilde{\mathbf{H}}_f\times\mathbf{l})-(\mathbf{l}\cdot\tilde{\mathbf{H}}_f)\tilde{\mathbf{H}}_f\label{Leq_l}\\
&&-\frac{\tilde{a}\gamma}{2}\Big(\alpha\dot{\mathbf{l}}+(\tilde{\beta}+\alpha)\tilde{u}\mathbf{l}'+\tilde{u}\tilde{c}_{\rm H}(1-\alpha\tilde{f}_{\rm so})\mathbf{l}\times\hat{\mathbf{y}}-\frac{\gamma}{2}\mathbf{H}_l\Big),\nonumber\\
&&\tilde{a}=\frac{4J_{\rm AF}}{M_{\rm s}a^3_0(1+\alpha^2)}, \tilde{\mathbf{H}}_f=\frac{\gamma \mathbf{H}_0}{1+\alpha^2}-\tilde{u}\tilde{c}_{\rm H}\tilde{f}_{\rm so}\hat{\mathbf{y}},\label{Hf}
\end{eqnarray}
with effective field $\mathbf{H}_{l}\equiv-(1/M_{\rm s}a^3_0)\delta W/\delta \mathbf{l}$, external out-of-plane magnetic field $\mathbf{H}_0=H_0\hat{\mathbf{z}}$, and we have defined the symbol of convective derivative as
\begin{eqnarray}
&&\mathcal{D}(...)\equiv (\partial_t+\tilde{u}\partial_x)(...),
\end{eqnarray}
with the spatial derivative component stemming from the adiabatic STT~\cite{Tatara00}. In this derivation, we only keep terms up to leading orders of $\mathbf{m}$ and Gilbert damping $\alpha$.
The Lagrangian density $\mathcal{L}$ and Rayleigh dissipation density $\mathcal{R}$ which can satisfy Eq.~(\ref{Leq_l}) via the Lagrange equation,
\begin{eqnarray}
\sum_{\mu=t,x}\partial_\mu\frac{\partial\mathcal{L}}{\partial (\partial_\mu\mathbf{l})}-\frac{\partial\mathcal{L}}{\partial\mathbf{l}}+\frac{\partial \mathcal{R}}{\partial\dot{\mathbf{l}}}=0,
\end{eqnarray}
can be found by using the procedure in~\cite{Gomonay2010} as [see Appendix B]
\begin{eqnarray}
\mathcal{L}&=&\rho (\mathcal{D}\mathbf{l})^2-2\rho (\tilde{\mathbf{H}}_f\cdot\mathbf{l}\times\mathcal{D}\mathbf{l})-\rho (\mathbf{l}\cdot\tilde{\mathbf{H}}_f)^2-\mathcal{W}(\mathbf{l}),\nonumber\\
\mathcal{R}&=&\frac{M_{\rm s}}{\gamma}\Big(\alpha\dot{\mathbf{l}}^2+2(\tilde{\beta}+\alpha)\tilde{u} \mathbf{l}'\cdot\dot{\mathbf{l}}+2\tilde{u}\tilde{c}_{\rm H}(1-\alpha\tilde{f}_{\rm so})\mathbf{l}\times\hat{\mathbf{y}}\cdot\dot{\mathbf{l}}\Big),\nonumber\\
\mathcal{W}&=&\frac{J_{\rm F}}{a_0}|\nabla\mathbf{l}|^2-\frac{2K_z}{a^3_0}(\mathbf{l}\cdot\hat{\mathbf{z}})^2+\mathcal{H}_{\rm D,N}+\mathcal{H}_{\rm D,B},\label{LR}
\end{eqnarray}
where $\rho=2M_{\rm s}/(\gamma^2\tilde{a})$, $J_{\rm F}$ is the ferromagnetic exchange coupling within each layer, $K_{z}$ is anisotropy energy constant for the easy-axis $\hat{\mathbf{z}}$, and we include both interfacial (we term it as N\'{e}el-type, $\mathcal{H}_{\rm D,N}$) and bulk (we term it as Bloch-type, $\mathcal{H}_{\rm D,B}$) DMI densities respectively as
\begin{eqnarray}
\mathcal{H}_{\rm D,N}&=&2D_{\rm N}\Big[l_z(\nabla\cdot\mathbf{l})-(\mathbf{l}\cdot\nabla)l_z\Big],\nonumber\\
\mathcal{H}_{\rm D,B}&=&2D_{\rm B}\Big[\hat{\mathbf{z}}\cdot\mathbf{l}\times \nabla l_z+l_z(\hat{\mathbf{z}}\cdot\nabla\times\mathbf{l})\Big].
\end{eqnarray}
In this work we take $D_{\rm N},D_{\rm B}>0$. The Lagrangian and Rayleigh dissipation we obtain in Eq.~(\ref{LR}) are the generalization of that derived by Gomonay and Loktev~\cite{Gomonay2010} by including STT and SOT, and this is the first central result of this work. Notably, the tilded parameters in Eq.~(\ref{LR}) directly reflect the effective torque strengths renormalized by the Gilbert damping. These Lagrangian and Rayleigh dissipation are generally applicable to any magnetic textures in synthetic AFMs, when the same kinds of torques are considered. The sublattice inequivalence of e.g., strengths of SOT, may be taken into account in a future work by generalizing the procedure of finding the Lagrangian in this work.
\subsection{II.2 Intuitive view of SkHE by SOT-induced AF exchange torque}
Before solving the Lagrange equations for the skyrmion dynamics, let us first give an intuitive view of how the dampling-like SOT drives the N\'{e}el (Bloch) bilayer skyrmions into longitudinal (transverse) motion relative to the current. 
In Fig.~\ref{Fig_intuitive}~(a--c) respectively, we schematically show the magnetizations of bilayer skyrmions in the N\'{e}el-type, Bloch-type, and in a general state with arbitrary helicity $\zeta$ when both interfacial and bulk DMIs coexist. The out-of-plane magnetic moments are indicated by blue and red colors, while the in-plane moments at the peripheries of static skyrmions are shown by black arrows. 
The SOT pointing in the direction of $\mathbf{M}\times\mathbf{M}\times\hat{\mathbf{y}}$ tends to rotate the local moments at periphery towards the $-\hat{\mathbf{y}}$ direction for the skyrmions in both layers, with the resultant tilted moments schematically shown by the blue arrows. However, after this small rotation, the local moments located on the same site in $xy$ plane in both layers are no longer perfectly antiparallel, and the interlayer AF exchange coupling will induce a strong exchange field $\mathbf{H}^{\rm B/A}_{\rm AF}\parallel -\mathbf{M}_{\rm A/B}$ in the opposite layer with the field directions shown by orange arrows. These fields will exert an exchange torque $\bm{\tau}^{\rm A/B}_{\rm AF}\parallel \mathbf{H}^{\rm A/B}_{\rm AF}\times\mathbf{M}_{\rm A/B}$ on the local moments at periphery denoted by squared blue dots and red crosses.
A similar consideration indicates that the center magnetic moment in the upper (lower) layer experiences an exchange torque pointing in the left (right) direction, irrespective of the helicity of the skyrmion magnetizations.
By focusing on the quasi-one-dimensional DW-like profiles inside the yellow dotted boxes, it is qualitatively expected that these torques will tend to move the N\'{e}el (Bloch) bilayer skyrmions into longitudinal (transverse) motion. For a pair of general bilayer skyrmions with helicity $\zeta$ as Fig.~\ref{Fig_intuitive}~(c) shows, the Hall angle is expected to be
\begin{eqnarray}
\theta_{\rm H}=-\zeta,
\end{eqnarray}
as found by using the Thiele equation in Ref.~\cite{Msiska}. This is a similar mechanism as the exchange torque in synthetic AFMs which can efficiently drive the DW motion at a speed as high as $\sim$1 km/s~\cite{Yang_Parkin,Yang_Parkin_2017}.
\begin{figure}[h]
\centering
\includegraphics[scale=0.36]{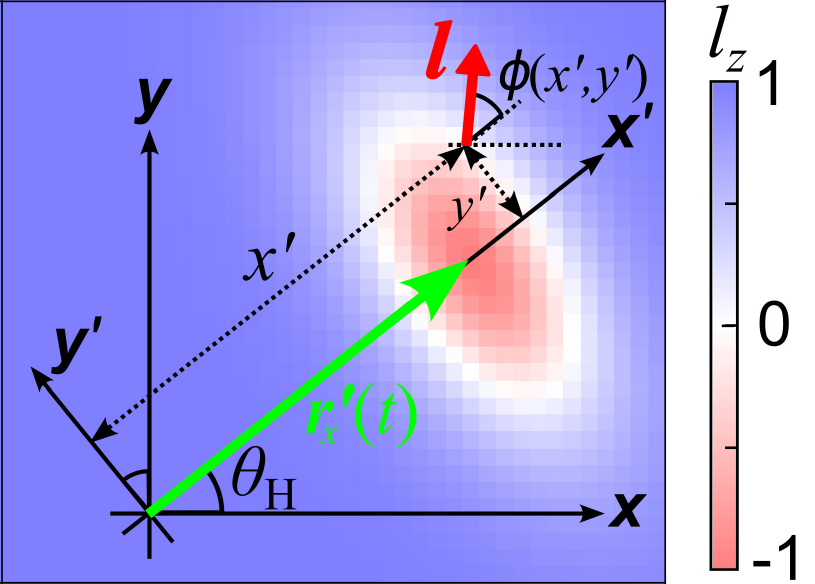}
\caption{Coordinate transformation with a presumed elliptical skyrmion profile and finite Hall angle $\theta_{\rm H}$. The color shows the out-of-plane components of $\mathbf{l}$ of bilayer skyrmions. A single in-plane $\mathbf{l}$ at the periphery is shown by a red arrow to indicate the angle $\phi(x',y')$ measured in the primed frame.}
\label{Fig_coordinate}
\end{figure}
\subsection{II.3 Lagrange equations for deformed bilayer skyrmions}
To verify the above intuitive expectation, we calculate the dynamics of bilayer skyrmions with different helicities by using the Lagrangian formalism. Based on the elliptical shape deformation found for bilayer N\'{e}el skyrmions~\cite{Salimath}, we consider a pair of antiferromagnetically coupled skyrmions with unknown Hall angle $\theta_{\rm H}$ and size parameters $\Delta_x, \Delta_y$ respectively along the longitudinal and transverse directions relative to the velocity, which are to be solved by Lagrange equation. 
Using the coordinate transformation depicted in Fig.~\ref{Fig_coordinate}, supposing that in the primed frame the skyrmion pair moves in $\hat{\mathbf{x}}'$ direction, with an elliptical shape carrying principal axes parallel to $\hat{\mathbf{x}}'$ and $\hat{\mathbf{y}}'$ directions, we define $\mathbf{l}=(\sin\theta\cos\phi,\sin\theta\sin\phi,\cos\theta)$ and take an ansatz for the angles generalized from Ref.~\cite{Salimath} as
\begin{eqnarray} &&\theta(x',y')=4\tan^{-1}(e^{\sqrt{X'^2+Y'^2}}),\nonumber\\
&&\phi(x',y')=n_{\rm v}\tan^{-1}\Big(\frac{Y'}{X'}\Big)+\zeta,\nonumber\\
&&X'=\frac{x'-r'_x(t)}{\Delta_x},\ Y'=\frac{y'}{\Delta_y}.\label{ansatz}
\end{eqnarray}
These angles are measured in the primed frame, with $n_{\rm v}$ and $\zeta$ being the vorticity and helicity of skyrmions, respectively. In this work we only consider the lowest-energy skyrmion with $n_{\rm v}=1$~\cite{Nagaosa}. The skyrmion center is assumed to be located at $r'_x(t)\hat{\mathbf{x}}'$ in the primed frame. The skyrmion radius $R_{x,y}$ along the $\hat{\mathbf{x}}',\hat{\mathbf{y}}'$ directions are defined by the periphery with 
\begin{eqnarray}
&&l_z(x'=r'_x\pm R_x,y'=0)=0,\\
&&l_z(x'=r'_x,y'=\pm R_y)=0, 
\end{eqnarray}
which give the relation
\begin{eqnarray}
R_{x,y}=\ln[\tan(3\pi/8)]\Delta_{x,y}\approx 0.88\Delta_{x,y},\label{Rxy}
\end{eqnarray}
from the definition of the ansatz. The radius contraction in $\hat{\mathbf{x}}'$ direction is assumed from the approximate Lorentz invariance in AFMs inspired by the well-known result for AF DWs~\cite{Shiino,Tatara,Gomonay2}. Since a radius contraction in $\hat{\mathbf{x}}'$ direction enhances the ferromagnetic exchange energy, it is physically expected that the radius in $\hat{\mathbf{y}}'$ direction should elongate to diminish this stored energy, as confirmed in the theory for N\'{e}el AF skyrmions by Salimath \textit{et al.}~\cite{Salimath}.

We substitute this skyrmion ansatz into the Lagrangian and Rayleigh dissipation densities and integrate over space [see Appendix C], and take the terminally steady state with $\dot{\Delta}_{x,y}=0$ and $\ddot{r}_{x,y}=0$ with $(r_x,r_y)=r'_{x}(\cos\theta_{\rm H},\sin\theta_{\rm H})$ measured in the original frame in Fig.~\ref{Fig_coordinate}. With dynamical variables taken as $r_x,r_y,\Delta_x$, and $\Delta_y$, after some algebra, we get four coupled Lagrange equations as
\begin{eqnarray}
v_x&=&\frac{\beta}{\alpha} u-\frac{C_1}{\alpha} uc_{\rm H}\Delta_x\cos\zeta,\label{vx}\\
v_y&=&\frac{C_1}{\alpha} uc_{\rm H}\Delta_x\sin\zeta,\label{vy}
\end{eqnarray}
and
\begin{widetext}
\begin{eqnarray}
0&=&-C_2\rho(\tilde{u}\tilde{c}_{\rm H}\tilde{f}_{\rm so})^2+C_3\Big(\rho \Big(\frac{\gamma{H}_0}{1+\alpha^2}\Big)^2-\frac{2{K}_z}{a^3_0}\Big)-C_1 D\Big(\frac{1}{\Delta_x}+\frac{1}{\Delta_y}\Big)\cos(\eta-\zeta)\nonumber\\
&+&(\tilde{u}\tilde{c}_{\rm H} \tilde{f}_{\rm so}) C_1 \rho \Big\{\frac{\tilde{u}}{v^2}\Big[\cos\zeta\Big(\frac{v^2_x}{\Delta_x}+\frac{v^2_y}{\Delta_y}\Big)-\sin\zeta~v_x v_y\Big(\frac{1}{\Delta_x}-\frac{1}{\Delta_y}\Big)\Big]-\frac{1}{\Delta_x}(v_x\cos\zeta-v_y\sin\zeta)\Big\},\label{L03}\\
0&=&\frac{{J}_{\rm F}}{a_0}\Big(\frac{1}{\Delta^2_x}-\frac{1}{\Delta^2_y}\Big)- \frac{\rho(v^2-2\tilde{u}v_x)}{\Delta^2_x}+\frac{\rho \tilde{u}^2}{v^2} \Big(\frac{v^2_y}{\Delta^2_y}-\frac{v^2_x}{\Delta^2_x}\Big)+C_1 D\Big(\frac{1}{\Delta_x}-\frac{1}{\Delta_y}\Big)\cos(\eta-\zeta)\nonumber\\
&+&(\tilde{u}\tilde{c}_{\rm H} \tilde{f}_{\rm so}) C_1 \rho \Big\{\frac{\tilde{u}}{v^2}\Big[\cos\zeta\Big(-\frac{v^2_x}{\Delta_x}+\frac{v^2_y}{\Delta_y}\Big)+\sin\zeta~v_x v_y\Big(\frac{1}{\Delta_x}+\frac{1}{\Delta_y}\Big)\Big]+\frac{1}{\Delta_x}(v_x\cos\zeta-v_y\sin\zeta)\Big\},\label{L04}
\end{eqnarray}
\end{widetext}
in which we have defined the DMI parameters $D$ and $\eta$ as
\begin{eqnarray}
&&D\equiv\sqrt{D^2_{\rm N}+D^2_{\rm B}},\ \eta\equiv \tan^{-1}(D_{\rm B}/D_{\rm N}),
\end{eqnarray}
and the constants $C_1\approx 0.7$, $C_2\approx 0.37$, and $C_3\approx 0.74$ are coming from integrals of the Lagrangian and Rayleigh dissipation densities over an infinite space as shown in Appendix C.

From Eq.~(\ref{vx}--\ref{vy}) which are expressed by the parameters in original LLGS equation [Eq.~(\ref{T})], we immediately get our first central conclusion. The N\'{e}el (Bloch) bilayer skyrmions in a synthetic AFM with $\zeta=0$ or $\pi~(\pi/2$ or $3\pi/2)$ have zero (finite) transverse velocity $v_y$, and this is directly induced by the dampling-like SOT since it is proportional to $c_{\rm H}$. This result quantitatively justifies our intuitive expectation of the motion driven by SOT in Fig.~\ref{Fig_intuitive}. Meanwhile, when considering the adiabatic limit with $\beta=0$ as in Ref.~\cite{Msiska}, Eq.~(\ref{vx}--\ref{vy}) give exactly their linear relation between Hall angle and helicity, namely,
\begin{eqnarray} 
\frac{v_y}{v_x}\equiv\tan\theta_{\rm H}=-\tan\zeta\ \ (\beta=0).
\end{eqnarray}
However, when nonadiabatic STT is finite there must be a nonzero longitudinal spin drift velocity component $\beta u/\alpha$ as in the case of current-driven DW motion~\cite{LeeDW}, regardless of the skyrmion helicity. More importantly, in this Lagrangian approach one can obtain the explicit relation between skyrmion velocity and shape deformation, and their dependences on current can be found by numerically solving the Lagrange equations in the next subsection.

Interestingly, we note that when $\beta=0$, from Eq.~(\ref{vx}) the N\'{e}el and Bloch bilayer skyrmions have $v_x=\pm \frac{C_1}{\alpha}uc_{\rm H}\Delta_x$ and $v_x=0$, respectively. This formula has exactly the same form (apart from numerical constants) as the velocity of an SOT-driven N\'{e}el/Bloch DW in intrinsic AFMs~\cite{Shiino} (see Eq.~(6) of~\cite{Shiino} for the N\'{e}el DW case, in which their DW width $\lambda$ is replaced by our skyrmion size parameter $\Delta_x$, and their parameter $B_D$ corresponds to our $c_{\rm H}$. The Bloch DW case has zero $v_x$ as stated in their text). This similarity may be understood if one approximately considers the DW as a one-dimensional horizontal extraction of the skyrmion texture near its periphery in a line passing through the skyrmion center, and restricts it to move in the $\pm\hat{\mathbf{x}}$ direction. 

To find possible analytical shapes of skyrmions in a limited regime, we note that typically the magnitude of field-like SOT is much smaller than damping-like SOT~\cite{Msiska}, thus here we assume $f_{\rm so}=0$. For simplicity, we also consider the case without external magnetic field, $H_0=0$. For any type of helicity, the interfacial and bulk DMIs are minimized by $\cos(\eta-\zeta)=-1$, and we approximately ignore the terms proportional to $(\tilde{u}\tilde{c}_{\rm H}\tilde{f}_{\rm so})\propto (\alpha u c_{\rm H})$ in Eq.~(\ref{L03}--\ref{L04}). Then these two equations have exact solutions of $\Delta_{x,y}$ as
\begin{eqnarray}
&&\Delta_x=\frac{\Delta_0}{2}\Big(1+\sqrt{1-F(\mathbf{v})}\Big),\nonumber\\ 
&&\Delta_y=\frac{\Delta_0}{2}\Big(1+\frac{1}{\sqrt{1-F(\mathbf{v})}}\Big),\nonumber\\
&&F(\mathbf{v})=\frac{[v^2-\tilde{u}(v_x+v_y)][v^2-\tilde{u}(v_x-v_y)]}{v^2_g v^2-\tilde{u}^2v^2_y},\label{F}
\end{eqnarray}
with the static skyrmion radius parameter $\Delta_0=\Delta_{x,y}(\tilde{u}=0)=\frac{a_0^3 C_1 D}{C_3 K_z}$ with $F(\mathbf{v})=0$ in the absence of current, and the magnon group velocity $v_g=\sqrt{(J_{\rm F}/a_0-\Delta_0C_1D/2)/\rho}$. For $0<F(\mathbf{v})<1$, these expressions clearly show the contraction and elongation of $\Delta_x$ and $\Delta_y$, respectively. When considering N\'{e}el bilayer skyrmions in the absence of STT, we take $\tilde{u}=0, v_y=0$ [from Eq.~(\ref{vy})] in Eq.~(\ref{F}), then it reduces to $F({v})=v^2/v^2_g$, which gives the familiar (inversed) Lorentz factor
\begin{eqnarray}
\sqrt{1-F}=\sqrt{1-v^2/v^2_g},
\end{eqnarray}
as found in~\cite{Salimath}. 
Interestingly, the above formulae can support real radius parameters $\Delta_{x,y}$ only when $F(\mathbf{v})<1$, thus it gives a criterion of instability of bilayer skyrmions as $F(\mathbf{v})\ge 1$. In our micromagnetic simulations with current density  $J\le 1.2\times10^{12}$~A/m$^2$, we found $F$ is always positive and smaller than 1, and we did not observe the instable skyrmions using our model parameters, but this criterion may be useful for future studies with different model parameters in other materials.
\subsection{II.4 Comparison with micromagnetic simulation}
To evaluate the robustness of this Lagrangian formalism, the four coupled Lagrange equations Eq.~(\ref{vx}--\ref{L04}) are solved numerically to compare with micromagnetic simulations.  We consider the case without external magnetic field ($H_0=0$) and perform simulations by numerically integrating the LLGS equation [Eq.~(\ref{LLGS1})] using the fourth-order Runge-Kutta algorithm. The energy in a discretized square lattice takes the form
\begin{eqnarray}
&&W=\sum_{i}\Big\{-\sum_{\delta=\hat{\mathbf{x}},\hat{\mathbf{y}}}J_{\rm F}\Big(\mathbf{M}_{\text{A},i}\cdot\mathbf{M}_{\text{A},i+\delta}+\mathbf{M}_{\text{B},i}\cdot\mathbf{M}_{\text{B},i+\delta}\Big)\nonumber\\
&&+J_{\rm AF}\mathbf{M}_{\text{A},i}\cdot\mathbf{M}_{\text{B},i}-K_{z}(M^2_{\text{A},iz}+M^2_{\text{B},iz})+w^{\rm B/N}_{\text{D},i}\Big\},
\end{eqnarray}
where $\mathbf{M}_{\text{A/B},i}$ is the magnetization direction vector in discretized space, and $w^{\rm B/N}_{\text{D},i}$ is either the bulk (Bloch) or interfacial (N\'{e}el) DMI respectively defined by
\begin{eqnarray}
&&w^{\rm B}_{\text{D},i}=\sum_{j=\rm A,B}a^2_0D_{\rm B}\Big({M}^z_{j,i}( {M}^y_{j,i+x} -{M}^x_{j,i+y})\nonumber\\
&&\hspace*{2.5cm}+{M}^x_{j,i} {M}^z_{j,i+y}-{M}^y_{j,i} {M}^z_{j,i+x}\Big),\nonumber\\
&&w^{\rm N}_{\text{D},i}=\sum_{j=\rm A,B}a^2_0D_{\rm N}\Big({M}^z_{j,i}( {M}^x_{j,i+x}+{M}^y_{j,i+y})\nonumber\\
&&\hspace*{2.5cm}-{M}^x_{j,i}{M}^z_{j,i+x}-{M}^y_{j,i}{M}^z_{j,i+y}\Big).
\end{eqnarray}
The details of simulation are described in Appendix D.

\begin{figure}[h]
\centering
\includegraphics[scale=0.4]{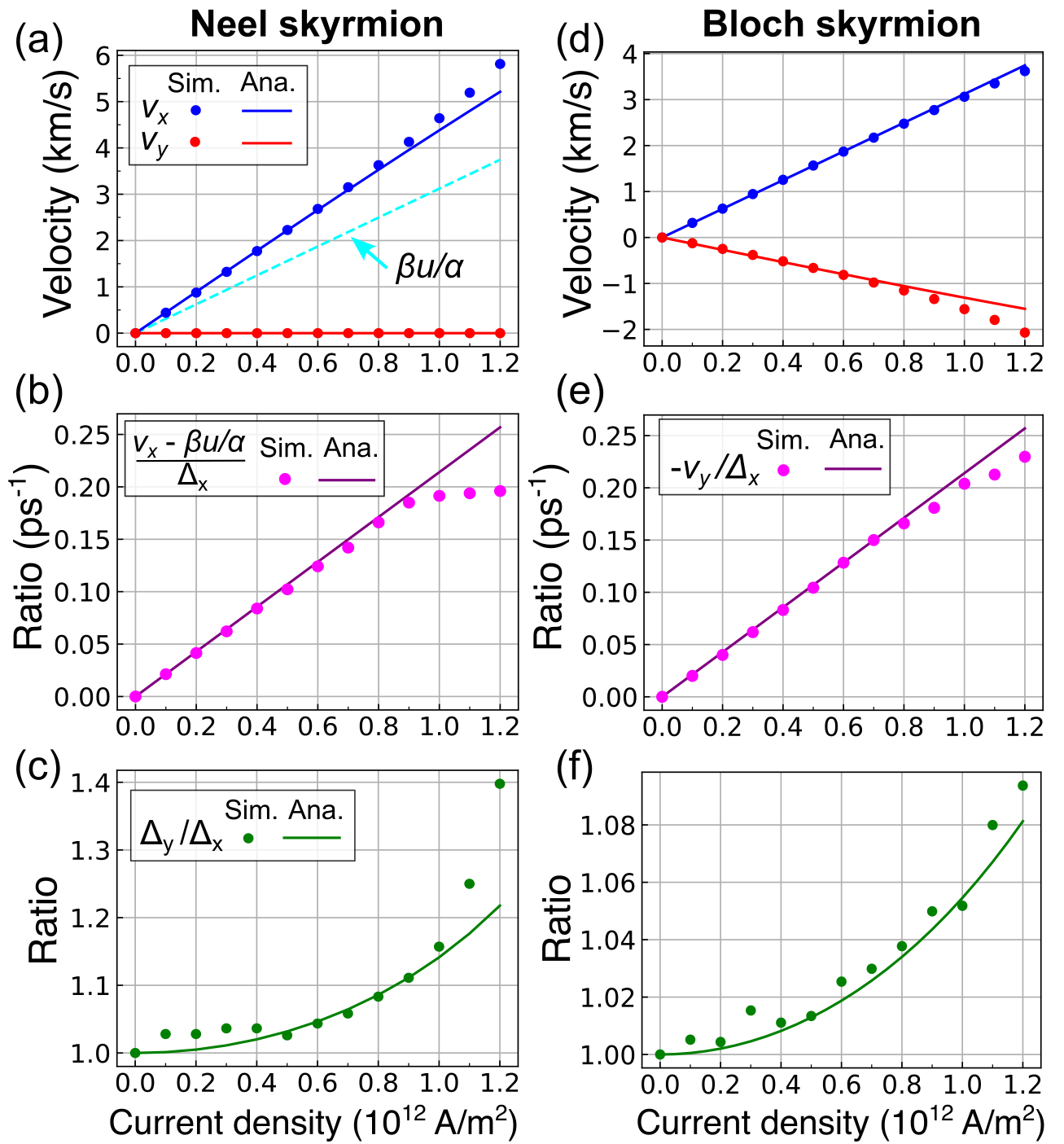}
\caption{Results of skyrmion velocity and shape parameters by solving Lagrange equations (curves) and by micromagnetic simulation (dots).}
\label{Fig_v}
\end{figure}
The model parameters are taken as $J_{\rm F}=a_0A_{\rm F}$ with $A_{\rm F}=13.18$~pJ/m and $a_0=1$~nm, $J_{\rm AF}=a_0A_{\rm AF}$ with $A_{\rm AF}=6.59$~pJ/m, $K_z/a^3_0=0.09$~MJ/m$^3$, $D=0.6$~mJ/m$^2$, $\alpha=0.05, \beta=2\alpha$, $M_{\rm s}=376$~kA/m, $\theta_{\rm sh}=0.5$, $t_{\rm L}=5a_0$, and $p=0.5$. We consider the case with only damping-like SOT ($f_{\rm so}=0$) in the main text. 
For N\'{e}el ($\eta=0$, $\zeta=\pi$) and Bloch ($\eta=\frac{\pi}{2}$, $\zeta=-\frac{\pi}{2}$) bilayer skyrmions as shown respectively in Fig.~\ref{Fig_v}~(a) and (d), the blue and red curves are solutions of velocity components $v_x$ and $v_y$ derived from numerically solving the Lagrange equations, which we denote as ``analytical (Ana.)", while the blue and red dots show the simulation (Sim.) results. It is clearly seen that for N\'{e}el (Bloch) bilayer skyrmions, a zero (finite) $v_y$ is found, meaning there is vanishing (finite) SkHE. The simulation results are in good agreement with analytics for low currents, but at large current densities near $1\times10^{12}$~A/m$^2$, there are small deviations, which may result from distortion of the skyrmion shape beyond a simple ellipse and/or spin wave emissions due to a large driving current.

Intriguingly, from Eq.~(\ref{vx}--\ref{vy}) there is a strong dependence of the velocity on longitudinal radius parameter $\Delta_x$ comparable with the case of DW motion discussed in the previous section. In the case of N\'{e}el and Bloch skyrmions, we get respectively $(v_x-\beta u/\alpha)/\Delta_x=C_1uc_{\rm H}/\alpha$ and $-v_y/\Delta_x=C_1 uc_{\rm H}/\alpha$, which are linear in the spin drift velocity $u\propto J$, with a slope proportional to the strength of damping-like SOT $c_{\rm H}$. In Fig.~\ref{Fig_v}~(b) and (e), we plot the results of these two functions versus the current density, and the approximately linear trends are also found in the simulation results, apart from the deviation in the large-current regime. This robust dependence may be observable, and it can provide a potential way to measure the strength of SOT in experiments. These plots show the overall consistency of the relation between skyrmion velocity and their shape parameter predicted by our Lagrangian formalism compared to simulations.

We note that the Lagrangian and Rayleigh dissipation found in this work can be generally applied to study the dynamics and/or deformations of any magnetic textures under the same kinds of STTs and SOTs. However, since the ansatz in Eq.~(\ref{ansatz}) is not the exact solution of skyrmions that minimizes the energy at static state, the solutions of Lagrange equations may be ansatz-dependent.
In particular, although the static skyrmion radius parameter $\Delta_0=\frac{a_0^3 C_1 D}{C_3 K_z}$ derived from this ansatz has the same proportionality to $D/K_z$ as found in~\cite{Salimath}, it does not depend on the intralayer ferromagnetic exchange coupling $J_{\rm F}$. If we recall the analytical skyrmion radius calculated by a more sophisticated 360$^\circ$~DW model~\cite{Wang} (which gives a good agreement of static skyrmion profile comparing the analytics by energy minimization and micromagnetic simulation), we can see that this ansatz is capable of matching the static simulated skyrmion radius only when considering the limit of $\frac{(\pi D a^2_0)^2}{16J_{\rm F}K_z}\ll 1$ (in which case the radius $R$ in Eq.~(13) of~\cite{Wang} becomes proportional to $D/K_z$ as found by our ansatz).

Moreover, when a large current is applied, in simulation we find that although the bilayer skyrmions show an elliptical shape, they also expand in size in both longitudinal and transverse directions relative to velocity, which is a similar behavior as found in~\cite{Liu} for ferromagnetic skyrmions. However, by using the simple ansatz in Eq.~(\ref{ansatz}) we can only get contracted $\Delta_x$ and elongated $\Delta_y$ as indicated by Eq.~(\ref{F}). The expansion reveals some imbalanced torque towards the radial directions of the skyrmions, which may be revealed by considering a more sophisticated generalization of the ferromagnetic 360$^\circ$~DW ansatz to AFMs, which is beyond our scope in the current work.

Due to the skyrmion expansion, Eq.~(\ref{F}) indicates that $\Delta_0$ should be better taken as a fitting parameter of the simulated skyrmion radius. Alternatively, to capture the essential feature of elliptical deformation while minimizing artificial effects caused by insufficiency of the ansatz, we consider the regime $\frac{(\pi D a^2_0)^2}{16J_{\rm F}K_z}\ll 1$ (with the chosen model parameters in simulation fulfilling this criterion), and try to minimize the effect of $\Delta_0$. Taking the ratio $\Delta_y/\Delta_x=1/\sqrt{1-F}$ first eliminates the overall $\Delta_0$ dependence. In the chosen regime, the $\Delta_0$ term in $v_g$ can be ignored and we have $v_g\approx\sqrt{J_{\rm F}/a_0\rho M^2_{\rm s}}$, and $\Delta_y/\Delta_x$ is approximately independent of $\Delta_0$ when $\beta u/\alpha\gg v_x-\beta u/\alpha=C_1uc_{\rm H}\Delta_x$ in the N\'{e}el skyrmion case, or when $|v_x|\gg |v_y|$ in the Bloch skyrmion case. 
It can be seen from Fig.~\ref{Fig_v}~(a) and (d) that the ratios $(\beta u/\alpha)/(v_x-\beta u/\alpha)$ and $|v_x|/|v_y|$ in the N\'{e}el and Bloch cases are both about~3, so it may moderately fulfill the required condition. In Fig.~\ref{Fig_v}~(c) and (f), we find qualitative agreements of $\Delta_y/\Delta_x$ between analytics and simulation. This ratio is quite small as $\Delta_y/\Delta_x<1.4$ for current density $J\le 1.2\times10^{12}$~A/m$^2$, but we find it depends on the model parameters, and in some cases the radius ratio can be as large as 3 with a substantial elliptical shape deformation. For a small finite field-like SOT with $f_{\rm so}=0.1$, the results are similar as above and we show the comparison between analytics and simulation in Supp. Fig.~\ref{Fig_S1} in the appendix.
\begin{figure*}
\centering
\includegraphics[scale=0.65]{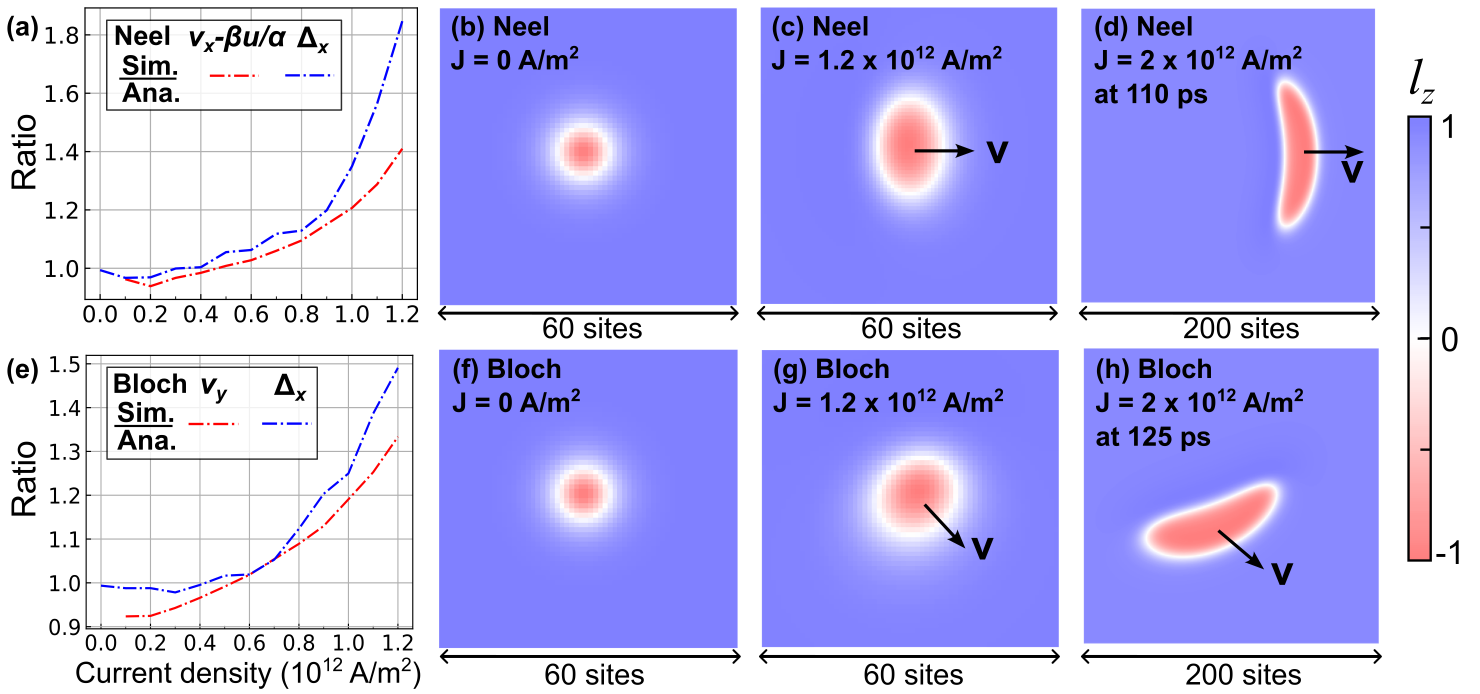}
\caption{The ratio between simulated and analytical results of $\Delta_x$-dependent parts of velocity (red dotted curves) and $\Delta_x$ (blue dotted curves) for (a) N\'{e}el and (e) Bloch bilayer skyrmions. (b)--(d) show the snapshots of simulated results of $l_z$ for N\'{e}el skyrmion case, while (f)--(h) are for Bloch skyrmions case. In (b),(f), the snapshots are taken for the relaxed skyrmion profiles before the current injection. In (c),(g), the snapshots are taken at 250 ps after the current injection. The black arrows indicate the velocity directions.}
\label{Fig_shape}
\end{figure*}
\section{III. Discussion}
The deviations from analytical velocities in Fig.~\ref{Fig_v}~(a),~(d) as well as the factors in Fig.~\ref{Fig_v}~(b),~(e) compared to simulations at large currents may be attributed to the skyrmion expansion, together with shape distortion beyond a simple ellipse and possible spin wave emissions. In Fig.~\ref{Fig_shape}~(a) and (e), we draw the ratio between the simulated and analytical results of the $\Delta_x$-dependent parts of velocity (red dotted curves) versus the ratio between the simulated and analytical $\Delta_x$ (blue dotted curves). The blue dotted curves show a clear expansion of simulated $\Delta_x$ at large currents compared with the analytical one. The trends of the red and blue dotted curves are roughly comparable, implying that the size expansion causes the deviation of velocity calculated by Lagrange equations. 
For a better illustration, the simulated skyrmion staggered magnetizations $l_z$ at zero current and at $J=1.2\times 10^{12}$ A/m$^2$ are plotted respectively in Fig.~\ref{Fig_shape}~(b),~(c) for N\'{e}el skyrmions and in (f),~(g) for Bloch skyrmions, which clearly show the elliptical shapes together with expansions driven by current.

When the current density becomes even larger as $2\times10^{12}$ A/m$^2$, we find the skyrmions cannot reach a rigidly moving state and their shapes will keep expanding in time until extending to the whole system with a length of 200 sites in each direction (with periodic boundary conditions). In Fig.~\ref{Fig_shape}~(d) and (h), we show the skyrmion profile snapshots at some instants, and find that the elliptical shape of skyrmions is further distorted into a crescent shape. Moreover, for the Bloch skyrmion case in Fig.~\ref{Fig_shape}~(h), the crescent shape does not contain a symmetry axis and the velocity direction indicated by the black arrow is also not parallel to any symmetry direction of the shape, in contrast to the N\'{e}el case in Fig.~\ref{Fig_shape}~(d) and to the lower current regimes in Fig.~\ref{Fig_shape}~(c),~(g). These properties indicate that the simple elliptical ansatz in Eq.~(\ref{ansatz}) is no longer reliable at large currents. However, since the scale of current density around $10^{12}$ A/m$^2$ is quite large, the Lagrangian formalism developed in this work with the elliptical ansatz should be robust at the low current regime which is more relevant for spintronics applications with lower power consumption.

As discussed in Sec.~II.3, the velocity of bilayer skyrmions we found in a synthetic AFM has a similar correspondence with that of DWs in intrinsic AFMs. But we note that in intrinsic AFMs the exchange interaction induces an additional parity breaking term proportional to $\mathbf{m}\cdot\mathbf{l}'$ in the Hamiltonian~\cite{Shiino,Tveten}, which lifts the degeneracy when switching the two sublattices (namely, taking $\mathbf{l}\rightarrow -\mathbf{l}$, then $\mathbf{m}\cdot\mathbf{l}'$ will change sign) for a magnetic texture. This term is absent in the case of synthetic AFMs since now the sublattice-exchange degeneracy exists when we consider the approximate situation without sublattice inequivalence of SOT, or DMI, etc. It seems that this parity-breaking term may be irrelevant to the skyrmion dynamics as the N\'{e}el skyrmion in a synthetic AFM here shows the similar form of longitudinal velocity as the N\'{e}el DW in intrinsic AFMs. An extension of our theory to intrinsic AF textures is left for future work.
\section{IV. Conclusion}
In this work, we have derived a Lagrangian formalism for synthetic AFMs based on the LLGS equation with the strengths of current-induced torques being renormalized by the Gilbert damping, which can be generally applied to any magnetic textures. The derivation can also be extended to consider different situations such as inequivalent SOT, DMI, or Gilbert dampings, etc., in the two sublattices. As an application, we use it to solve the dynamics of bilayer skyrmions by taking a simplified skyrmion profile ansatz, with the analytical results by numerically soving Lagrange equations matching well with mircromagnetic simulations. Importantly, we find the zero (nonzero) skyrmion Hall effect for N\'{e}el (Bloch) bilayer skyrmions driven by SOT, and we find an approximate elliptical shape deformation with strong dependence of the skyrmion velocity on the longitudinal radius parallel to velocity. Although the overall expansion of skyrmions cannot be captured by this simple ansatz, we believe an extension of the more realistic 360$^\circ$ DW model to AFMs can be used to get the accurate quantitative results. These findings are important for the applications of AF skyrmions to spintronics such as the skyrmion racetrack memory by using synthetic AFMs.
\section{Acknowledgements}
This work is supported by JSPS KAKENHI (No.~20H00337, No.~24H02231, and No.~25H00611), JST CREST (No.~JPMJCR20T1), and Waseda University Grant for Special Research Project (Grant No.~2024C-153, No.~2025C-133, and No.~2025C-134). M.-K.L. is grateful for illuminating discussions with Collins Ashu Akosa, Xichao Zhang, and Rintaro Eto.
\section{Appendices}
\subsection{A. Derivation of a closed equation of motion for the staggered magnetization}\label{AppA}
From the LLGS equation for the two sublattices in Eq.~(\ref{LLGS1}), using the definitions $\mathbf{m}=(\mathbf{M}_{\rm A}+\mathbf{M}_{\rm B})/2$ and $\mathbf{l}=(\mathbf{M}_{\rm A}-\mathbf{M}_{\rm B})/2$, we can derive the corresponding exact LLGS equations for $\mathbf{m}$ and $\mathbf{l}$ as
\begin{widetext}
\begin{eqnarray}
\mathcal{D}\mathbf{m}&=&\frac{\gamma}{2}\Big(\tilde{\mathbf{H}}_m\times\mathbf{m}+\tilde{\mathbf{H}}_l\times \mathbf{l}\Big)-\frac{\alpha\gamma}{2}\Big(
\mathbf{m}\times\mathbf{m}\times\tilde{\mathbf{H}}_m
+\mathbf{m}\times\mathbf{l}\times\tilde{\mathbf{H}}_l
+\mathbf{l}\times\mathbf{m}\times\tilde{\mathbf{H}}_l
+\mathbf{l}\times\mathbf{l}\times\tilde{\mathbf{H}}_m\Big)\nonumber\\
&+&\tilde{\beta} \tilde{u}\Big(\mathbf{m}\times\mathbf{m}'+\mathbf{l}\times\mathbf{l}'\Big)+\tilde{u}\tilde{f}_{\rm so}\tilde{c}_{\rm H} (\mathbf{m}\times\hat{\mathbf{y}})+\tilde{u}\tilde{c}_{\rm H}\Big[\mathbf{m}\times(\mathbf{m}\times\hat{\mathbf{y}})+\mathbf{l}\times(\mathbf{l}\times\hat{\mathbf{y}})\Big],\\
\mathcal{D}\mathbf{l}&=&\frac{\gamma}{2}\Big(\tilde{\mathbf{H}}_m\times\mathbf{l}+\tilde{\mathbf{H}}_l\times \mathbf{m}\Big)-\frac{\alpha\gamma}{2}\Big(
\mathbf{m}\times\mathbf{m}\times\tilde{\mathbf{H}}_l
+\mathbf{m}\times\mathbf{l}\times\tilde{\mathbf{H}}_m
+\mathbf{l}\times\mathbf{m}\times\tilde{\mathbf{H}}_m
+\mathbf{l}\times\mathbf{l}\times\tilde{\mathbf{H}}_l\Big)\nonumber\\
&+&\tilde{\beta} \tilde{u}\Big(\mathbf{m}\times\mathbf{l}'+\mathbf{l}\times\mathbf{m}'\Big)+\tilde{u}\tilde{f}_{\rm so}\tilde{c}_{\rm H} (\mathbf{l}\times\hat{\mathbf{y}})+\tilde{u}\tilde{c}_{\rm H}\Big[\mathbf{m}\times(\mathbf{l}\times\hat{\mathbf{y}})+\mathbf{l}\times(\mathbf{m}\times\hat{\mathbf{y}})\Big],
\end{eqnarray}
\end{widetext}
with $\tilde{\mathbf{H}}_{m,l}=\mathbf{H}_{m,l}/(1+\alpha^2)$. We approximate the right-hand sides by keeping only terms up to the first order of $\mathbf{m}$ below. 
After summation of the energy over the two sublattices and assuming $J_{\rm AF}$ is the largest energy scale in the exchange limit, the dominant energy for $\mathbf{m}$ is ${W}_{\mathbf{m}}\approx \sum_i( 2J_{\rm AF} \mathbf{m}^2_i-2M_{\rm s}a^3_0\mathbf{H}_0\cdot\mathbf{m}_i)$~\cite{Gomonay2010} with $\mathbf{m}_i$ labeling the discrete averaged magnetization unit vector at site $i$, and the summation of $i$ is over the square lattice. The effective field for $\mathbf{m}_i$ is thus approximately $\tilde{\mathbf{H}}_m=\frac{-1}{1+\alpha^2}\frac{\partial W_{\mathbf{m}}}{M_{\rm s}a^3_0\partial\mathbf{m}_i}\approx\tilde{\mathbf{H}}_{0}-\tilde{a}\mathbf{m}_i$ with $\tilde{\mathbf{H}}_{0}=\frac{2\mathbf{H}_0}{1+\alpha^2}$ and $\tilde{a}=\frac{4J_{\rm AF}}{M_{\rm s}a^3_0(1+\alpha^2)}$. Note that $\tilde{a}\sim$ Tesla since $\mathbf{m}_i$ is dimensionless. Then these LLGS equations can be written as
\begin{widetext}
\begin{eqnarray}
\mathcal{D}\mathbf{m}&\approx&\frac{\gamma}{2}\Big(\tilde{\mathbf{H}}_0\times\mathbf{m}+\tilde{\mathbf{H}}_l\times \mathbf{l}\Big)
-\frac{\alpha\gamma}{2}\Big(\mathbf{m}\times\mathbf{l}\times\tilde{\mathbf{H}}_l
+\mathbf{l}\times\mathbf{m}\times\tilde{\mathbf{H}}_l
+\mathbf{l}\times\mathbf{l}\times(\tilde{\mathbf{H}}_0-\tilde{a}\mathbf{m})\Big)\nonumber\\
&+&\tilde{\beta} \tilde{u}(\mathbf{l}\times\mathbf{l}')+\tilde{u}\tilde{f}_{\rm so}\tilde{c}_{\rm H} (\mathbf{m}\times\hat{\mathbf{y}})+\tilde{u}\tilde{c}_{\rm H}[\mathbf{l}\times(\mathbf{l}\times\hat{\mathbf{y}})]\label{Dm},\\
\mathcal{D}\mathbf{l}&\approx&\frac{\gamma}{2}\Big(\tilde{\mathbf{H}}_0\times\mathbf{l}-\tilde{a}\mathbf{m}\times\mathbf{l}+\tilde{\mathbf{H}}_l\times \mathbf{m}\Big)-\frac{\alpha\gamma}{2}\Big(
\mathbf{m}\times\mathbf{l}\times\tilde{\mathbf{H}}_0
+\mathbf{l}\times\mathbf{m}\times\tilde{\mathbf{H}}_0
+\mathbf{l}\times\mathbf{l}\times\tilde{\mathbf{H}}_l\Big)\nonumber\\
&+&\tilde{\beta} \tilde{u}\Big(\mathbf{m}\times\mathbf{l}'+\mathbf{l}\times\mathbf{m}'\Big)+\tilde{u}\tilde{f}_{\rm so}\tilde{c}_{\rm H} (\mathbf{l}\times\hat{\mathbf{y}})+\tilde{u}\tilde{c}_{\rm H}\Big[\mathbf{m}\times(\mathbf{l}\times\hat{\mathbf{y}})+\mathbf{l}\times(\mathbf{m}\times\hat{\mathbf{y}})\Big].
\end{eqnarray}
\end{widetext}
Consider $\frac{2}{\tilde{a}\gamma}\mathbf{l}\times\mathcal{D}\mathbf{l}$ by using the second equation, we can get an expression for $\mathbf{m}$ as
\begin{eqnarray}
\mathbf{m}&=&\frac{1}{\tilde{a}}\Big\{\tilde{\mathbf{H}}_0-(\mathbf{l}\cdot\tilde{\mathbf{H}}_0)\mathbf{l}-(\mathbf{l}\cdot\tilde{\mathbf{H}}_l) \mathbf{m}\\
&&-\alpha\Big(
(\mathbf{l}\cdot\mathbf{m}\times\tilde{\mathbf{H}}_0)\mathbf{l}-\mathbf{m}\times\tilde{\mathbf{H}}_0
-\mathbf{l}\times\tilde{\mathbf{H}}_l\Big)\nonumber\\
&&+\frac{2\tilde{\beta} \tilde{u}}{\gamma}(\mathbf{l}\times\mathbf{l}\times\mathbf{m}')+\frac{2\tilde{u}\tilde{f}_{\rm so}\tilde{c}_{\rm H}}{\gamma} (\mathbf{l}\times\mathbf{l}\times\hat{\mathbf{y}})\nonumber\\
&&+\frac{2\tilde{u}\tilde{c}_{\rm H}}{\gamma}\Big[(\mathbf{l}\cdot\mathbf{m}\times\hat{\mathbf{y}})\mathbf{l}-(\mathbf{m}\times\hat{\mathbf{y}})\Big]+\frac{2}{\gamma}\mathcal{D}\mathbf{l}\times\mathbf{l}\Big\}.\nonumber
\end{eqnarray}
Since the overall denominator $\tilde{a}\propto J_{\rm AF}$, in the exchange limit, on the right-hand side we ignore terms linear in $\mathbf{m}$, then it can be simplified as
\begin{eqnarray}
\mathbf{m}&\approx&\frac{2}{\tilde{a}\gamma}\Big\{\mathcal{D}\mathbf{l}\times\mathbf{l}-\frac{\alpha\gamma}{2}\tilde{\mathbf{H}}_l\times\mathbf{l}-(\mathbf{l}\times\mathbf{l}\times\tilde{\mathbf{H}}_f)\Big\},\label{m}
\end{eqnarray}
with $\tilde{\mathbf{H}}_f=\gamma\tilde{\mathbf{H}}_0/2-\tilde{u}\tilde{c}_{\rm H}\tilde{f}_{\rm so}\hat{\mathbf{y}}$. This equation shows that $\mathbf{m}$ is a slave variable that can be expressed solely by the dynamics of $\mathbf{l}$. It is clear that $\mathbf{m}$ is small due to the large $\tilde{a}\propto J_{\rm AF}$ in the overall denominator. The requirement of $\mathbf{m}\cdot\mathbf{l}=0$ by the definitions of $\mathbf{m}$ and $\mathbf{l}$ is also fulfilled by this form of $\mathbf{m}$. Next, we try to eliminate the appearance of $\mathbf{m}$ to find a closed equation of motion for $\mathbf{l}$ only. First, we take the convective derivative of $\mathbf{m}$ by using Eq.~(\ref{m}) and compare it with Eq.~(\ref{Dm}) to get
\begin{widetext}
\begin{eqnarray}
\mathcal{D}\mathbf{m}&=&\frac{2}{\tilde{a}\gamma}\Big\{\mathcal{D}^2\mathbf{l}\times\mathbf{l}-\frac{\alpha\gamma}{2}\mathcal{D}\tilde{\mathbf{H}}_l\times\mathbf{l}
-\frac{\alpha\gamma}{2}\tilde{\mathbf{H}}_l\times\mathcal{D}\mathbf{l}
-(\mathcal{D}\mathbf{l}\cdot\tilde{\mathbf{H}}_f)\mathbf{l}-(\mathbf{l}\cdot\tilde{\mathbf{H}}_f)\mathcal{D}\mathbf{l}-\mathbf{l}\times(\mathbf{l}\times\mathcal{D}\tilde{\mathbf{H}}_f)\Big\}\\
&=&-\mathbf{m}\times\tilde{\mathbf{H}}_f
-\frac{\alpha\gamma}{2}\Big((\mathbf{m}\cdot\tilde{\mathbf{H}}_l)\mathbf{l}+(\mathbf{l}\cdot\tilde{\mathbf{H}}_l)\mathbf{m}
+\tilde{a}\mathbf{m}\Big)+\frac{\gamma}{2}\tilde{\mathbf{H}}_l\times \mathbf{l}
-\frac{\alpha\gamma}{2}(\mathbf{l}\times\mathbf{l}\times\tilde{\mathbf{H}}_0)+\tilde{\beta} \tilde{u}(\mathbf{l}\times\mathbf{l}')+\tilde{u}\tilde{c}_{\rm H}(\mathbf{l}\times\mathbf{l}\times\hat{\mathbf{y}}).\nonumber
\end{eqnarray}
\end{widetext}
By substituting Eq.~(\ref{m}) for $\mathbf{m}$ in the second line, after some tedious algebra we get a closed equation of motion for $\mathbf{l}$ as (we have used definition $(1+\alpha)\tilde{\mathbf{H}}_{l}=\mathbf{H}_{l}$)
\begin{widetext}
\begin{eqnarray}
\mathcal{D}^2\mathbf{l}\times\mathbf{l}&=&\Big[2(\tilde{\mathbf{H}}_f\times\mathcal{D}\mathbf{l})+(\mathcal{D}\tilde{\mathbf{H}}_f\times\mathbf{l})-(\mathbf{l}\cdot\tilde{\mathbf{H}}_f)\tilde{\mathbf{H}}_f\Big]\times\mathbf{l}-\frac{\tilde{a}\gamma}{2}\Big\{\alpha\dot{\mathbf{l}}
+(\tilde{\beta}+\alpha)\tilde{u}\mathbf{l}'
+\tilde{u}\tilde{c}_{\rm H}(1-\alpha\tilde{f}_{\rm so})\mathbf{l}\times\hat{\mathbf{y}}-\frac{\gamma}{2}{\mathbf{H}}_l\Big\}\times\mathbf{l}\nonumber\\
&-&\frac{\alpha\gamma}{2}\Big\{(\mathbf{l}\cdot\tilde{\mathbf{H}}_f)\tilde{\mathbf{H}}_l+(\mathbf{l}\cdot\tilde{\mathbf{H}}_l)\tilde{\mathbf{H}}_f+(\mathcal{D}\mathbf{l}\times\mathbf{l}\cdot\tilde{\mathbf{H}}_l)\mathbf{l}
-2(\mathbf{l}\cdot\tilde{\mathbf{H}}_f)(\mathbf{l}\cdot\tilde{\mathbf{H}}_l)\mathbf{l}+(\mathbf{l}\cdot\tilde{\mathbf{H}}_l)(\mathcal{D}\mathbf{l}\times\mathbf{l})\nonumber
\\
&&\hspace*{0.5cm}-\mathcal{D}\tilde{\mathbf{H}}_l\times\mathbf{l}
-\tilde{\mathbf{H}}_l\times\mathcal{D}\mathbf{l}-\frac{\alpha\gamma}{2}(\mathbf{l}\cdot\tilde{\mathbf{H}}_l)(\tilde{\mathbf{H}}_l\times\mathbf{l})\Big\}.
\end{eqnarray}
\end{widetext}
The dominant terms are in the first line. The terms in the second and third lines are proportional to $\alpha$ compared to the squared bracket in the first line, and they do not carry the factor $\tilde{a}$ compared to the curly bracket in the first line, therefore, we ignore them for simplicity. Then the equation becomes
\begin{eqnarray}
&&\Big[(\mathcal{D}^2\mathbf{l})-\Big\{2(\tilde{\mathbf{H}}_f\times\mathcal{D}\mathbf{l})+(\mathcal{D}\tilde{\mathbf{H}}_f\times\mathbf{l})-(\mathbf{l}\cdot\tilde{\mathbf{H}}_f)\tilde{\mathbf{H}}_f\nonumber\\
&&-\frac{\tilde{a}\gamma}{2}\Big(\alpha\dot{\mathbf{l}}+(\tilde{\beta}+\alpha)\tilde{u}\mathbf{l}'+\tilde{u}\tilde{c}_{\rm H}(1-\alpha\tilde{f}_{\rm so})\mathbf{l}\times\hat{\mathbf{y}}-\frac{\gamma}{2}\mathbf{H}_l\Big)\Big\}\Big]\nonumber\\
&&\times\mathbf{l}=0\label{LLGS_l}.
\end{eqnarray}
A special solution of $\mathbf{l}$ satisfies the restricted equation that $\mathcal{D}^2\mathbf{l}$ equals the summation of the terms in the curved bracket on the right-hand side, as Eq.~(\ref{Leq_l}) in the main text. The Lagrange equation by taking the Lagrangian in Eq.~(\ref{LR}) and the Rayleigh dissipation found in the next section can be shown to satisfy Eq.~(\ref{LLGS_l}).
\subsection{B. Derivation of the Rayleigh dissipation function}
We derive the Rayleigh dissipation function by considering the time derivative of energy following the procedure in~\cite{Gomonay2010}. The rate of energy dissipation for sublattice $j(=\rm A,B)$, $\frac{dW_j}{dt}=\frac{\partial W}{\partial \mathbf{M}_j}\cdot\dot{\mathbf{M}}_j$, can be written in the continuous limit as (we omit sublattice index $j$ for convenience)
\begin{eqnarray}
&&\frac{dW}{dt}=-M_{\rm s}a_0(1+\alpha^2)\int d^2r\tilde{\mathbf{H}}^{\rm eff}\cdot\dot{\mathbf{M}}\label{dwdt0}\\
&&=M_{\rm s}a_0(1+\alpha^2)\int d^2r\Big\{\tilde{\mathbf{H}}^{\rm eff}\cdot \tilde{u}\mathbf{M}'-\tilde{\mathbf{H}}^{\rm eff}\times\mathbf{M}\cdot[...]\Big\},\nonumber
\end{eqnarray}
with a shorthand notation $[...]\equiv[-\alpha\gamma\mathbf{M}\times\tilde{\mathbf{H}}^{\rm eff}+\tilde{\beta} \tilde{u}\mathbf{M}'+\tilde{u}\tilde{c}_{\rm H} \mathbf{M}\times\hat{\mathbf{y}}+\tilde{u}\tilde{f}_{\rm so}\tilde{c}_{\rm H}\hat{\mathbf{y}}]$. In the second line we have invoked the LLGS equation Eq.~(\ref{LLGS1}) to replace $\dot{\mathbf{M}}$ in the first line. Using Eq.~(\ref{LLGS1}) again, the second integrand in the second line can be reduced to
\begin{eqnarray}
&&-\tilde{\mathbf{H}}^{\rm eff}\times\mathbf{M}\cdot[...]=\frac{-1}{\gamma}\mathcal{D}\mathbf{M}\cdot[...].\label{integrand1}
\end{eqnarray}
On the other hand, the first integrand can be transformed into a convenient form by using Eq.~(\ref{LLGS1}) to get
\begin{eqnarray}
\mathbf{M}\cdot \mathbf{M}'\times \mathcal{D}\mathbf{M}&=&-\gamma \mathbf{M}'\cdot\tilde{\mathbf{H}}^{\rm eff}+\mathbf{M}'\cdot[...]\nonumber\\
&=&\mathbf{M}\cdot \mathbf{M}'\times \dot{\mathbf{M}},
\end{eqnarray}
such that the first integrand becomes
\begin{eqnarray}
\tilde{u}\mathbf{M}'\cdot\tilde{\mathbf{H}}^{\rm eff}=\frac{\tilde{u}}{\gamma}\mathbf{M}'\cdot[...]-\frac{\tilde{u}}{\gamma}\dot{\mathbf{M}}\cdot\mathbf{M}\times\mathbf{M}'.\label{integrand2}
\end{eqnarray}
Adding Eq.~(\ref{integrand1}) and Eq.~(\ref{integrand2}), Eq.~(\ref{dwdt0}) gives
\begin{eqnarray}
&&\frac{dW}{dt}=\frac{-M_{\rm s}a_0(1+\alpha^2)}{\gamma}\int d^2r\dot{\mathbf{M}}\cdot\Big\{\tilde{u}\mathbf{M}\times\mathbf{M}'\label{dwdt}\\
&&+\tilde{\beta} \tilde{u}\mathbf{M}'-\alpha\gamma\mathbf{M}\times\tilde{\mathbf{H}}^{\rm eff}+\tilde{u}\tilde{c}_{\rm H} \mathbf{M}\times \hat{\mathbf{y}}+\tilde{u}\tilde{f}_{\rm so}\tilde{c}_{\rm H}\hat{\mathbf{y}}\Big\}.\nonumber
\end{eqnarray}
The $\mathbf{M}\times\tilde{\mathbf{H}}^{\rm eff}$ term in the integrand can be further reduced by LLGS equation [Eq.~(\ref{LLGS1})]. Due to the tilded parameters, it is not apparent to see the $\alpha$ dependences. After replacing all tilded parameters by the un-tilded ones, expanding up to the first order of $\alpha$, and adding both sublattices, the energy dissipation in the zeroth order of $\mathbf{m}$ can be written as~\cite{Gomonay2010}
\begin{eqnarray}
&&\sum_{j=\rm A,B}\frac{dW_j}{dt}\equiv -a_0\int \dot{\mathbf{l}}\cdot\frac{\partial \mathcal{R}}{\partial\dot{\mathbf{l}}}\ d^2r\nonumber\\
&&\approx\frac{-2M_{\rm s}a_0}{\gamma}\int d^2r\Big(\alpha\dot{\mathbf{l}}^2+(\tilde{\beta}+\alpha)\tilde{u}\mathbf{l}'\cdot \dot{\mathbf{l}}\nonumber\\
&&\hspace*{2cm}+\tilde{u}\tilde{c}_{\rm H}(1-\alpha\tilde{f}_{\rm so})\mathbf{l}\times\hat{\mathbf{y}}\cdot\dot{\mathbf{l}}\Big)
\end{eqnarray}
from which we obtain the Rayleigh dissipation function as shown in Eq.~(\ref{LR}) in the main text. Intriguingly, when being expressed by the original un-tilded parameters, the density of energy dissipation rate can be written as
\begin{eqnarray}
&&\frac{-2M_{\rm s}a_0}{\gamma}\Big(\alpha\dot{\mathbf{l}}^2+\beta u\mathbf{l}'\cdot \dot{\mathbf{l}}+uc_{\rm H}\mathbf{l}\times\hat{\mathbf{y}}\cdot\dot{\mathbf{l}}\Big),
\end{eqnarray}
in which the three terms sequentially stem from the Gilbert damping, nonadiabatic STT, and damping-like SOT, as expected. Note that in this equation the parameters $u,\beta$, and $c_{\rm H}$ are all expressed by the original parameters. Therefore, the renormalization effect on them due to the Gilbert damping seems to only occur in the energy dissipation in second or higher orders of $\alpha$, when we ignore the dissipation due to the averaged magnetization $\mathbf{m}$ as assumed above.
\subsection{C. Lagrangian terms in detail}
We integrate the Lagrangian and Rayleigh dissipation densities over all 2D space, $\int dxdy\mathcal{L}=\int dx'dy'\mathcal{L}=\Delta_x\Delta_y\int dX'dY'\mathcal{L}=\Delta_x\Delta_y\int^{\infty}_0Z dZ\int^{2\pi}_0d\psi\mathcal{L}\equiv\int\mathcal{L}$, with $Z\equiv \sqrt{X'^2+Y'^2}$ and $\psi\equiv\tan^{-1}(Y'/X')$. After integration, the Lagrangian per unit thickness contains following $L_i$ terms,
\begin{widetext}
\begin{eqnarray}
L_1&=&\rho\int (\mathcal{D}\mathbf{l})^2=\rho (I_1+I_2)\Big[\frac{\Delta_y}{\Delta_x}v^2+\Big(\frac{\Delta_y}{\Delta_x}\frac{v^2_x}{v^2}+\frac{\Delta_x}{\Delta_y}\frac{v^2_y}{v^2}\Big)\tilde{u}^2 -2\frac{\Delta_y}{\Delta_x}\tilde{u} v_x\Big],\label{L_terms}\\
L_2&=&-2\rho\int  \tilde{\mathbf{H}}_f\cdot\mathbf{l}\times\mathcal{D}\mathbf{l}\nonumber\\
&=&2\rho \tilde{u}\tilde{c}_{\rm H}\tilde{f}_{\rm so} (I_4+I_5)\Big[\Delta_y(v_y\sin\zeta-v_x\cos\zeta)+\frac{\tilde{u}}{v^2}\Big([v^2_y\Delta_x+v^2_x\Delta_y]\cos\zeta+v_xv_y(\Delta_x-\Delta_y)\sin\zeta\Big)\Big],\nonumber\\
L_3&=&\int\Big(-\rho (\mathbf{l}\cdot\tilde{\mathbf{H}}_f)^2+\frac{{2}{K}_z}{a^3_0}l^2_z\Big)\rightarrow-\Big(\frac{{2}{K}_z}{a^3_0}-\rho\Big(\frac{\gamma H_0}{1+\alpha^2}\Big)^2\Big)I_6\Delta_x\Delta_y-\rho (\tilde{u}\tilde{c}_{\rm H}\tilde{f}_{\rm so})^2I_3\Delta_x\Delta_y,\nonumber\\
L_4&=&\frac{-{J}_{\rm F}}{a_0}\int|\nabla\mathbf{l}|^2=\frac{-{J}_{\rm F}}{{a_0}}\Big(\frac{\Delta^2_x+\Delta^2_y}{\Delta_x\Delta_y}\Big)\Big(I_1+ I_2\Big),\nonumber\\
L_5&=&-\int(\mathcal{H}_{\rm D,N}+\mathcal{H}_{\rm D,B})=-{2}(I_4+I_5)(\Delta_x+\Delta_y)({D}_{\rm N}\cos\zeta+{D}_{\rm B}\sin\zeta).\nonumber
\end{eqnarray}
\end{widetext}
Since the integral $\int l^2_z=\int \cos^2\theta$ diverges, in $L_3$ we have replaced the integral from $\int l^2_z$ to $\int(l^2_z-1)$. Here we have included both interfacial ($\propto D_{\rm N}$) and bulk ($\propto D_{\rm B}$) DMIs. By the same integration procedure, the Rayleigh dissipation per unit thickness contains following $R_i$ terms
\begin{widetext}
\begin{eqnarray}
R_1&=&\frac{M_{\rm s}\alpha}{\gamma }\int \dot{\mathbf{l}}^2=\frac{M_{\rm s}\alpha}{\gamma }\frac{\Delta_y}{\Delta_x}v^2(I_1+I_2),\ \ R_2=\frac{2M_{\rm s}(\tilde{\beta}+\alpha)\tilde{u}}{\gamma}\int \mathbf{l}'\cdot\dot{\mathbf{l}}=\frac{-2M_{\rm s}(\tilde{\beta}+\alpha)\tilde{u} }{\gamma}\frac{\Delta_y}{\Delta_x}v_x(I_1+I_2),\nonumber\\
R_3&=&\frac{2M_{\rm s}\tilde{u}\tilde{c}_{\rm H}(1-\alpha\tilde{f}_{\rm so})}{\gamma }\int\mathbf{l}\times\hat{\mathbf{y}}\cdot\dot{\mathbf{l}}=\frac{2M_{\rm s}\tilde{u}\tilde{c}_{\rm H}(1-\alpha\tilde{f}_{\rm so})}{\gamma}\Delta_y(v_x\cos\zeta-v_y\sin\zeta)(I_4+I_5).\label{R_terms}
\end{eqnarray}
\end{widetext}
The constant integrals $I_j$ are as follows.
\begin{widetext}
\begin{eqnarray}
&&I_1=\int^{\infty}_0 RdR\int^{2\pi}_0d\psi\Big(\frac{2~\text{sech}R~X'}{R}\Big)^2=4\pi\ln 2,\ \ I_2=\int^{\infty}_0 RdR\int^{2\pi}_0d\psi\sin^2\theta\Big(\frac{Y'}{R^2}\Big)^2\approx 4.84,\nonumber\\
&&I_3=\int^{\infty}_0 RdR\int^{2\pi}_0d\psi\sin^2\theta\frac{X'^2}{R^2}\approx 5,\ \ I_4=\int^{\infty}_0 RdR\int^{2\pi}_0d\psi\frac{2~\text{sech}R~X'^2}{R^2}\approx 11.5,\nonumber\\
&&I_5=\int^{\infty}_0 RdR\int^{2\pi}_0d\psi\sin\theta\cos\theta\frac{X'^2}{R^3}\approx -2.09,\ \ I_6=\int^{\infty}_0 RdR\int^{2\pi}_0d\psi\sin^2\theta\approx 10,\nonumber
\end{eqnarray}
\end{widetext}
and we have defined $C_1=\frac{I_4+I_5}{I_1+I_2}\approx 0.7$, $C_2=\frac{I_3}{I_1+I_2}\approx 0.37$, and $C_3=\frac{I_6}{I_1+I_2}\approx 0.74$.

In detail, the following derivatives and functions of $\theta$ and $\phi$ have been used to derive $L_i$ and $R_i$,
\begin{widetext}
\begin{eqnarray}
&&\partial_t\theta= \frac{-2~\text{sech}Z~X'\dot{r}'}{Z\Delta_x},\ \partial_x\theta=\frac{2~\text{sech}Z}{Z}\Big[\frac{X'\cos\theta_{\rm H}}{\Delta_x}-\frac{Y'\sin\theta_{\rm H}}{\Delta_y}\Big],\ \partial_y\theta=\frac{2~\text{sech}Z}{Z}\Big[\frac{X'\sin\theta_{\rm H}}{\Delta_x}+\frac{Y'\cos\theta_{\rm H}}{\Delta_y}\Big],\nonumber\\
&&\partial_t\phi=\frac{Y'\dot{r}'}{Z^2\Delta_x},\ \partial_x\phi=\frac{-1}{Z^2}\Big[\frac{ X'\sin\theta_{\rm H}}{\Delta_y}+\frac{Y'\cos\theta_{\rm H}}{\Delta_x}\Big],\ \partial_y\phi=\frac{1}{Z^2}\Big[\frac{X'\cos\theta_{\rm H}}{\Delta_y}-\frac{Y'\sin\theta_{\rm H}}{\Delta_x}\Big],\nonumber\\
&&\sin\theta=-2~\text{sech}Z\tanh Z,\ \cos\theta=1-2~\text{sech}^2Z.\nonumber
\end{eqnarray}
\end{widetext}
\subsection{D. Details of micromagnetic simulation}\label{Sim}
We take a two-layer system stacked in the $z$ direction, with each layer containing 200 sites in both $x$ and $y$ directions. Periodic boundary conditions are used for magnetizations in the $xy$ plane. We first generate a static bilayer skyrmion texture without any applied current or field. After relaxation by the LLG equation (Eq.~(\ref{LLGS1}) in the main text without the current-induced torques), we simulate the skyrmion dynamics via LLGS equation, in which we increase the current from zero to the target value linearly in a period of 100~ps, then we keep it fixed at the target value. The fourth-order Runge-Kutta algorithm is used to numerically integrate both LLG and LLGS equations. The position of skyrmion center, which is assumed to be the site inside the skyrmion at which $|M_{\text{A/B},iz}|$ has the largest value, is recorded from 100~ps to 250~ps to calculate the averaged velocity in this period. 
To find the simulated skyrmion radius, we draw the skyrmion periphery in the final state near 250~ps by finding the points with the smallest $|M_{\text{A/B},iz}|$, then use the calculated velocity in simulation to draw four lines, two of which being with the slope of $v_y/v_x$ and the other two of $-v_x/v_y$, and tune the intercepts at which these four lines cross the $x, y$ axes to fit the skyrmion periphery. Differences of positions of these intercepts are used to calculate the skyrmion radius $R_x$ and $R_y$ respectively in $x'$ and $y'$ directions by using the slopes. The simulated values of $\Delta_{x,y}$ is extracted by using Eq.~(\ref{Rxy}) in the main text. We use homemade Julia and Python codes respectively for the integration of LLG/LLGS equation and the data analysis described above.
\begin{figure}[h]
\centering
\includegraphics[scale=0.4]{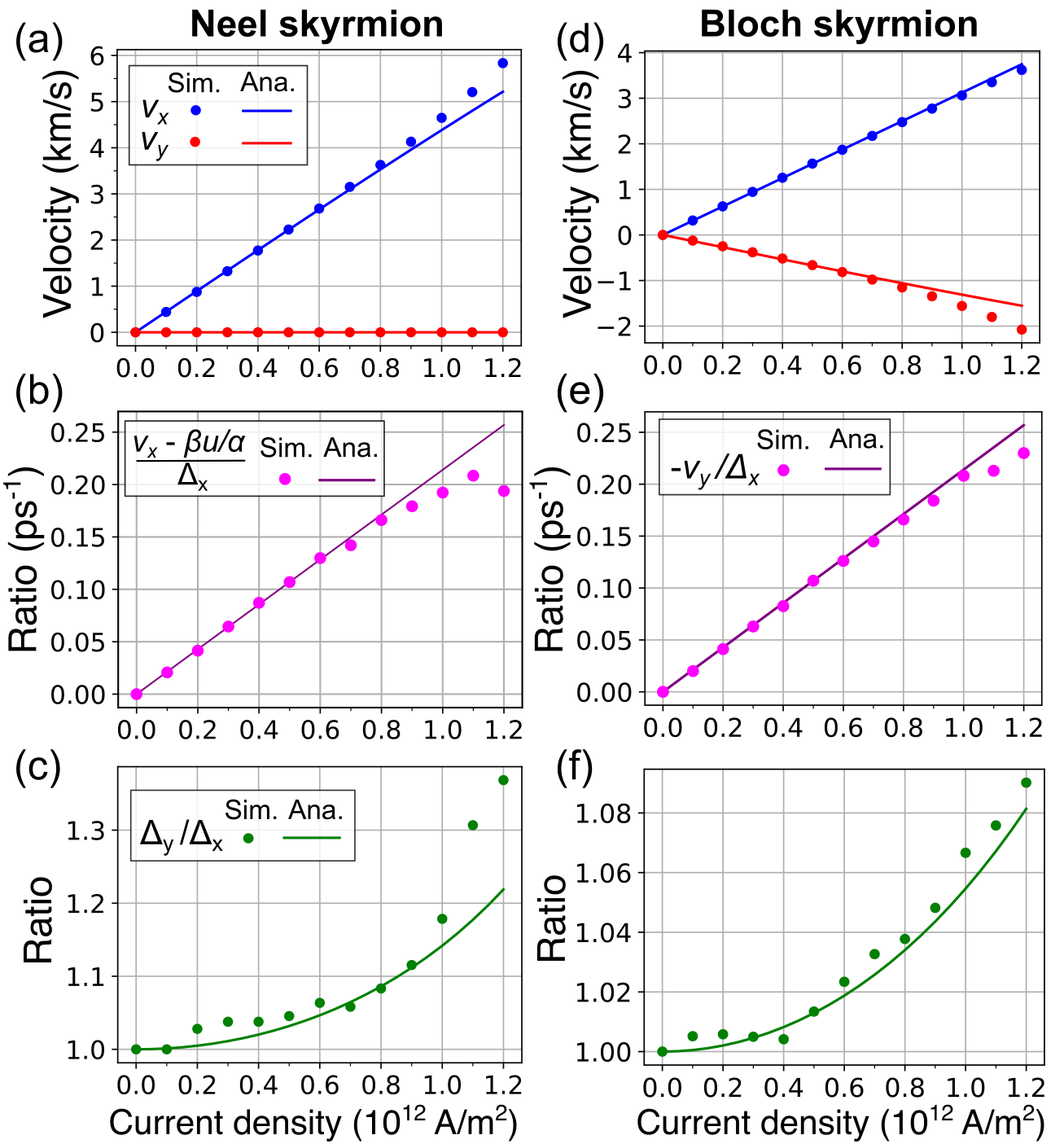}
\renewcommand{\figurename}{Supplementary Fig.}
\setcounter{figure}{0}
\caption{Results of skyrmion velocity and shape parameters by solving Lagrange equations (curves) and by micromagnetic simulation (dots) in the case of $f_{\rm so}=0.1$.}
\label{Fig_S1}
\end{figure}

\end{document}